\documentclass[a4paper]{VisionStyle}
\usepackage{epsfig}

\begin{document}

\title{The XMM-Newton Science Archive}

\author{C.\,Arviset \and M.\,Guainazzi \and J.\,Hernandez
  \and J.\,Dowson \and P.\,Osuna \and A.\,Venet } 

\institute{
  ESA Science Operations and Data Systems Division, 
  Research and Scientific Support Department, VILSPA, 
  Villafranca del Castillo Satellite Tracking Station,
  PoBox 50727, 28080 Madrid, Spain}

\maketitle 

\begin{abstract}

The XMM-NEWTON Science Archive (XSA) is being developed within the XMM-NEWTON Science Operations Centre at the ESA VILSPA Satellite Tracking Station of Villafranca del Castillo, near Madrid in Spain. The first public release is planned for April 2002.

Based on the ISO Data Archive architecture and design, the XSA will be accessible from the World Wide Web through a Java interface at http://xmm.vilspa.esa.es/xsa that will allow:
\begin{itemize}
  \item powerful and complex queries against the observations and exposure catalogue for the first release and against the source catalogue in future releases
  \item configurable results display, including product visualization tools
  \item customisable product retrieval via a shopping basket
  \item selection of product level
  \item product retrieval via FTP or CDROM
  \item proprietary rights protection for both display and data request
\end{itemize}

The XSA will also be ready to provide inter-operability with other archives and applications.

\end{abstract}

\section{Introduction}
The XMM-NEWTON data products are currently delivered by mean of CDROM to the observation Principle Investigators (PIs) using internal systems of the Science Operations Centre in VILSPA.

With the first proprietary data becoming public soon, there arose the need of a more general and user friendly data access and distribution system. 

In early 2001, it was decided to build a new system called the XMM-NEWTON Science Archive (XSA) to allow general astronomers easy access to available XMM-NEWTON data products.

A set of representative users from the SOC, the SSC (Survey Science Centre, who are responsible for the generation of XMM-NEWTON data products) and some external X-ray scientists met in March 2001 to define the user requirements for the XSA.
Matteo Guainazzi was nominated XSA Archive Scientist. His responsibility is to gather all users' feedback and requests and to consolidate them into the User Requirement Document (URD). He defines the implementation priorities for the software development team. Interacting daily with the developers, he makes sure that the XSA is developed according to the users' expectations.
The first issue of XSA URD was released in early April 2001 for the design and development of the archive to start in late April 2001. The goal was to have a public release ready as soon as the first proprietary data becomes public, i.e. April 2002.

\begin{figure}[ht]
  \begin{center}
{\tt see Fig1.gif}
  \end{center}
\caption{ XSA Query Panel }  
\label{carviset-WA3_fig:fig1}
\end{figure}

In order to be able to meet such a short deadline, it was decided to adopt the open and flexible JAVA-based architecture of the successful ISO Data Archive (IDA, available at  http://www.iso.vilspa.esa.es/ida) which would offer a fast and cost-efficient development while ensuring that all X-ray community specific user requirements are fulfilled.

\section{XSA General Presentation}
\subsection{Accessing the XSA}
The XSA contains all the XMM-NEWTON science data products and relevant auxiliary data as soon as they become available to the SOC. 

All the data products (ODF and PPS) are stored on hard disks with an  approximate overall planned volume of 2 TeraBytes. All relevant auxiliary meta data are stored in a database for the user to be able to query against, as well as on standard file for quick display. When the data are put into the XSA (first the ODF and later on the PPS products once processed), the observation PI is notified by e-mail about the availability of his data for him to access directly. The standard mechanism of data distribution on CDROM will still be available, but the XSA will also provide the PI with a faster (FTP) access to his data.

From early April 2002, the XSA can be accessed on the web at http://xmm.vilspa.esa.es/xsa. Its user interface is a Java applet that requires a Java 1.1 compatible web browser (such as Netscape 4.5 or Microsoft Internet Explorer 4 or higher). In case of browser incompatibility, the XSA applet can also be downloaded on the user machine and run as a standalone application for a variety of platforms (SUN Solaris, Digital Unix, Linux, Windows, Mac OS). The applet is digitally certified by the European Space Agency through a third party well established company Verisign. On his approval, the user can allow the XSA applet to access some of his local resources (e.g. printing or direct file download).

\subsection{Querying the XSA}
Through an intuitive interface, the user can perform complex queries at observations and exposure level against the XSA database (see Figure~\ref{carviset-WA3_fig:fig1}). The queries can be made from various panels that can be opened of closed to simplify the interface:
\begin{itemize}
  \item general astronomical criteria (observation id, target name (to be resolved by NED, SIMBAD or as given by the observer), target coordinates, list of targets, observation availability (public data, proprietary data for the observation PI), observation date and time, observations status, etc...)
  \item revolution and observation parameters (revolution number, phase in orbit, Galactic N H, observation mode)
  \item exposure parameters (exposure id, instrument, exposure duration, data mode, instrument filter)
  \item proposal information (proposal id, PI name, proposal category and proposal program)
\end{itemize}
  
\subsection{Observations and Exposure Catalogue}
When a query is made, the corresponding list of observations and exposures are returned together with relevant information and PPS icons (see Figure~\ref{carviset-WA3_fig:fig2}). The list of observations is returned by default, but it can be expanded to the list of exposures for each observation. In the future, it will be possible to access individual spectra and light curves for each source in the EPIC cameras field-of-view, detected by the automatic SSC reduction pipeline. The user can select the amount of information in the returned catalogue. Furthermore, one can choose how many items should be displayed. Navigation buttons (Start / End of List, Next / Previous) allows the user to access all elements in the returned list.

\begin{figure}[ht]
  \begin{center}
{\tt see Fig2.gif}
  \end{center}
\caption{ XSA Result Panel }  
\label{carviset-WA3_fig:fig2}
\end{figure}

The date on which the observation becomes public is indicated on the interface. In case the data is still proprietary, the PPS icon will not be displayed. The access to the PPS image, normally available by a simple click on the icon as indicated in Figure~\ref{carviset-WA3_fig:fig3}, will not be allowed either.

Login is not required to perform queries and view the result. However, once logged in, the user can select independently observations and exposures and move them to the shopping basket for later retrieval. More searches can be made to add further items to the shopping basket. Proprietary observations and exposure can only be moved to the shopping basket by the observation PI. Similarly, once logged in, observation PI will also see the PPS icon and PPS image instead of the default display "proprietary" flag.

\subsection{Retrieving Data Products}
In the shopping basket panel, the user can select the level of products he or she wants to retrieve for individual items, either the ODF or the PPS products or both. Once the request is submitted, the request ID is immediately returned.
Products are prepared by the XSA retrieval system and copied on the FTP area (public area if public data, and secure area if proprietary data). An e-mail is then sent to the user to let him or her know the request has been completed and that he or she can retrieve his/her data from the FTP area. A typical request of a few observations is generally processed by the system in less than 10 minutes, from the time of the initial request to the dispatch of the e-mail.

As an alternative, users can also request the data products to be put on CDROMs in order to have them mailed to them.

A quota is set for each user in order to avoid data retrieval traffic congestion and ensure to all users the quickest and most efficient access to the archive. The quota is currently based on a number of observations (10 maximum) and exposures (100 maximum) that can be retrieved by a user. The quota is reset every three days. Quota increase can be obtained by contacting the XMM-NEWTON helpdesk (xmmhelp@xmm.vilspa.esa.es). In the future, the quota may be changed to a retrieved data volume basis.

In addition to this powerful retrieval mechanism, a button on the observation and exposure catalogue also allows the user to download directly all data products (ODF and PPS) associated with that particular observation / exposure. This is saved directly as a compressed tar file on the user disk.

\begin{figure}[ht]
  \begin{center}
{\tt see Fig3.gif} \\
{\tt see Fig4.gif}
  \end{center}
\caption{ Examples of PPS products icons available in the XSA user interface: {\it Left panel}: EPIC field-of-view image; {\it Right panel}: RGS net source spectra }  
\label{carviset-WA3_fig:fig3}
\end{figure}

\subsection{Other Access to XMM-NEWTON Data}
As well as this standard access to the data, it is also planned to provide direct access to the PPS image file or data products from external archives or applications, bypassing the standard Java user interface. This will allow XMM-NEWTON data to be widely spread around commonly used X-ray tools.

\section{Open Architecture and Design}
The XSA is using an open 3-tier architecture design (see Figure ~\ref{carviset-WA3_fig:fig4}), similar to the one of the IDA. The main aim of it is to separate the data and the business logic from their presentation on the user interface. Such an approach will make the software maintenance easier and will permit to adapt the interface to newer technologies as they come. This will not require a full change of the database and data products organization.

\begin{figure}[ht]
  \begin{center}
{\tt see Fig5.gif}
  \end{center}
\caption{ XSA 3-tier Architecture }  
\label{carviset-WA3_fig:fig4}
\end{figure}

In general, Java provided the multi-platform support, having the same code for all operating systems, realising the dream of many developers: {\it write once, run anywhere}.
One can now go one step further by using XML to achieve multi-project support. In conjunction with the multi-platform capability of Java, the use of configurable XML files allows the same code to be used for various archives to reach {\it write once, use anywhere} capability.
Altogether, an open architecture with the use of Java and XML allows re-usability from one project to another, ensuring that the archive will be accessible on all user platforms. This brings significant cost savings in the development, test and maintenance phases.

\section{Future Plans}
There will be continuous improvements made to the XSA throughout the XMM-NEWTON operations at the rate of a major release every year. In between, there will be minor releases to fix the encountered problems and gradually bring in new improvements.

Feedback from the users' community is welcome, and shall be conveyed through the XMM-Newton HelpDesk (xmmhelp@xmm.vilspa.esa.es). By collecting and consolidating the users' inputs, defining implementation priorities, and interacting daily with the developers, the Archive Scientist will ensure that the XSA is delivered according to the scientific users expectations.

Currently, the major improvements planned for the coming versions are :
\begin{itemize}
  \item customized retrieval menu for PPS products ({\it i.e.}, a user will be able to retrieve whatever sub-set of PPS products he/she wants)
  \item source specific catalogue (when the corresponding SSC products will be routinely produced)
  \item "on-the-fly" filtering for events list
  \item creation of updated scientific products through XSA driven SAS reduction pipeline runs
  \item links to publications
  \item XID products (when available)
\end{itemize}

Moreover, discussions are also taking place to provide external archives and applications with the XMM-NEWTON observations/exposures log, together with a mechanism (via Java servlets) to remotely access the PPS images or even the full data products directly. In addition, the XSA should also be able to offer links to other astronomical archives to provide added-value to the user trying to retrieve XMM-NEWTON data. These inter-operability capabilities will become even more relevant in the context of the 
Virtual Observatory projects being currently developed.

\section{Conclusion}
The XMM-NEWTON Science Archive developed at ESA provides a new, easy and fast access to all available XMM-NEWTON data products and auxiliary data through a state-of-the art powerful Java interface available from April 2002 at http://xmm.vilspa.esa.esa/xsa. It allows all the X-ray astronomers to have quick access to the data and exploit the science observed by XMM-NEWTON.

After the fully functional first public release in April 2002, the XSA will continue to be improved to offer more capabilities and access to newer data products. Work will also be done to ensure the XSA's full integration into the Virtual Observatory projects.

\end{document}